\documentclass[10pt]{elsart}

\usepackage{amssymb,amsmath}
\usepackage{graphics,graphicx}
\usepackage{epsfig,epic,eepic,psfig,rotate,float}

\begin{document}
\runauthor{Freire and Gallas}
\begin{frontmatter}
\title{Spontaneous Emergence of Spatio-Temporal Order in
       Class 4 Automata}
\author[a1,a2]{Jason A.C.~Gallas\corauthref{cor}}       %\thanksref{jg}}
\corauth[cor]{Corresponding author: jgallas@if.ufrgs.br}
\author[a2]{Hans J.~Herrmann}  %\thanksref{hjh}}
\address[a1]{Instituto de F\'\i sica, 
             Universidade Federal do Rio Grande do Sul, \\
            91501-970 Porto Alegre, Brazil}
\address[a2]{Institut f\"ur Computerphysik, 
             Universit\"at Stuttgart, \\
             D-70569 Stuttgart, Germany}
%\thanks[hjh]{E-mail: hans@icp.uni-stuttgart.de}
%\thanks[jg]{Home-page: {\tt http://www.if.ufrgs.br/$\sim$jgallas} }

\begin{abstract}
We report surprisingly regular behaviors observed for
a class 4 cellular automaton, the totalistic rule 20:
starting from disordered initial configurations the
automaton produces patterns which are periodic not only 
in time but also in space.
This is the first evidence that different types of
spatio-temporal order can emerge under specific conditions
out of disorder in the same discrete rule based algorithm.
\end{abstract}
%\begin{keyword}
% aaa; aaa 
%\end{keyword}
\end{frontmatter}

%====================================================================
%===================== Introducao ===================================
%\section{Introduction}

Cellular automata (CA) are discrete dynamical systems
of many degrees of freedom that have been used successfully to describe
a plethora of nonlinear phenomena ranging from pattern
formation to chaos, have
been applied for modelization in biology, geology, chemistry, sociology,
etc and have even been proposed as minimal computer. 
In a seminal and remarkably influential work
Wolfram \cite{wolf84,ANKS} distinguished four classes of
behavior for {\it binary\/}
CA, i.e.~those with a local degree of freedom
$\sigma_i$ which is  either $\sigma_i=1$ or $\sigma_i=0$.
In class 1 all initial configurations evolve to a fixed
point, when all cells are either one or zero. 
Automata of class 2 evolve to spatially static structures
with eventually short-lived temporal oscillations. 
Class 3 automata produce self-similar patterns
that can be characterized as discrete chaos. 
Finally, the most interesting class 4 contains all 
remaining automata which,  according to Wolfram, display
behaviors  ``more complex  than chaos" 
being possibly capable of universal computation.
The famous Game of Life is a CA of class 4 \cite{mc90}. 

Wolfram has recently evoked class 4 automata as paradigms for the emergence
of life\cite{ANKS}. 
Our findings support this picture in the sense that
specific initial conditions in confined systems yield highly
non-trivial spatio-temporal periodic patterns while the overall
picture is dominated by long transients.
In fact, 
based on the behavior of the class 4 rule 20 automaton it has been 
argued \cite{gh90} that in the thermodynamic limit, 
the limit of infinite lattice size, class 4 automata are just having
extremely long transients but ending invariably in one of
the three simpler classes.

%==================================================
\begin{figure}
\unitlength 1mm
\begin{center}
\leavevmode 
{\includegraphics[width=6.87cm]{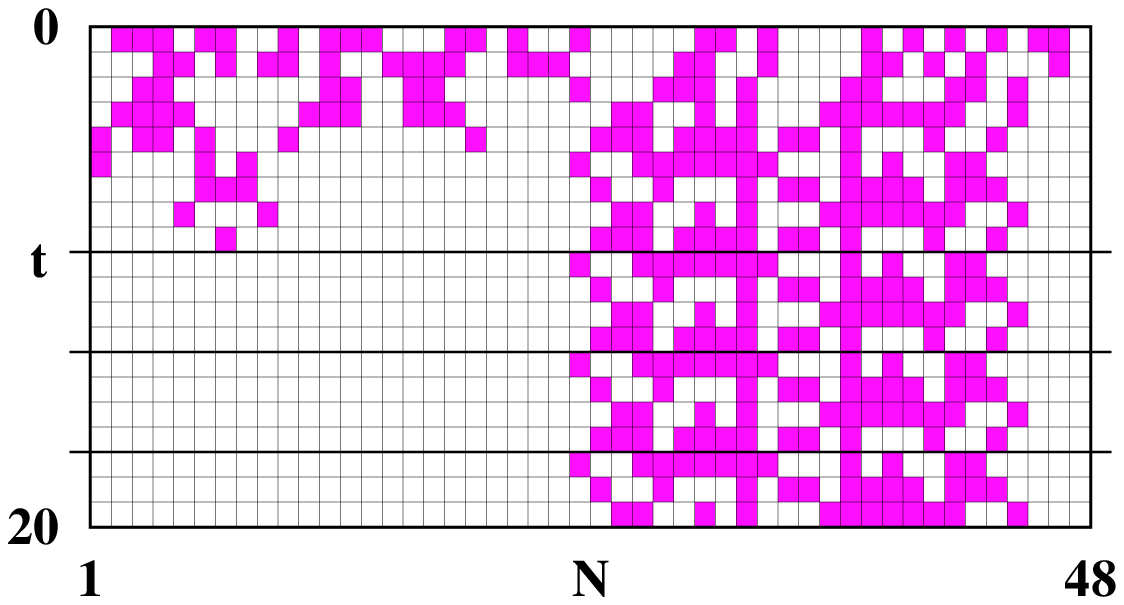}
\includegraphics[width=6.87cm]{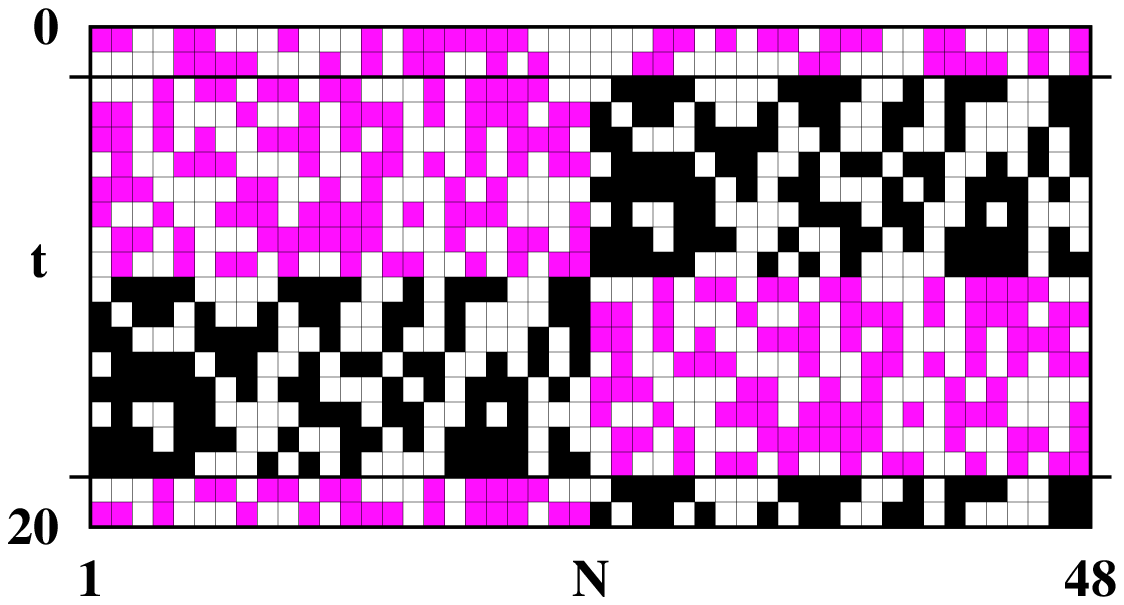}}%
%\vspace{-0.4truecm}}
\end{center}
\caption{\protect Left: a typical period-$4$ persistent
  structure (``glider'') emerging after a transient.
        Right: example of a remarkably regular space-filling pattern of
        period $16$. Two tonalities are used to represent
         ``1'' to better display inner symmetries.}
\label{fig:fig1}       
\end{figure}
%==================================================

The purpose of this short note is to illustrate graphically
a number of very regular behaviors that we observed 
and are studying on  class 4 automata. 
All results here are for rule 20 although similar
behavior exists for other rules.

The remarkable new behavior observed is that
starting from disordered initial configurations one finds
patterns which are periodic not only 
in time but also in space. We refer to such behavior as
{\it spatio-temporal\/} order.
We believe this to be the first evidence that different types of
spatio-temporal order can emerge under specific conditions
out of disorder in the same discrete rule based algorithm.
It also shows that although dependent on the initial condition, 
class 4 automata can yield {\it highly regular\/}
attractors which have not been observed before, not even for
the three simpler classes. 
This behavior is not to be confused with trivial static backgrounds
familiar from, e.g., nearest-neighbor 1D rule 110.
Interestingly, spatio-temporal order seems to predominate
for smaller systems since preliminary data suggest
that the probability of finding then may vanish in the thermodynamic limit,
thereby not excluding the speculation about intrinsic transient
behavior in class 4 automata \cite{gh90}.

Our cellular automaton is a linear chain of $N$ sites with
the aforementioned binary degree of freedom $\sigma_i$ and
periodic boundary conditions.
The $\sigma_i$ values define the {\it state\/} of each
site at a given time. 
For rule 20
the state of any given site $i$ at time $t+1$ depends on the 
state of the site at time $t$ as well as on the state of its
nearest and next-nearest neighbors through the sum
$\Sigma \equiv \sigma_{i-2}(t) +\sigma_{i-1}(t)
                +\sigma_i(t) +\sigma_{i+1}(t) +\sigma_{i+2}(t)$
and is synchronously updated  as follows:
%%%%%%%%%%%%%%%%%%%%%%%%%
%\begin{equation}
%\sigma_i(t+1)=1\ \hbox{  if  }\ \Sigma=2\hbox{ or }4, \qquad
%\sigma_i(t+1)=0\ \hbox{  otherwise}.
%\end{equation}
%%%%%%%%%%%%%%%%%%%%%%%%
\begin{equation}
 \sigma_i(t+1) = \begin{cases}
      \quad  1 & \hbox{ if  }\quad  \Sigma = 2 \hbox{ or } 4 \,, \\
      \quad  0 & \hbox{ otherwise}\,.
       \end{cases}   \label{rule20}
\end{equation}

%==================================================
\begin{figure}[thb]
\unitlength 1mm
\begin{center}
\leavevmode 
{\includegraphics[width=4.75cm]{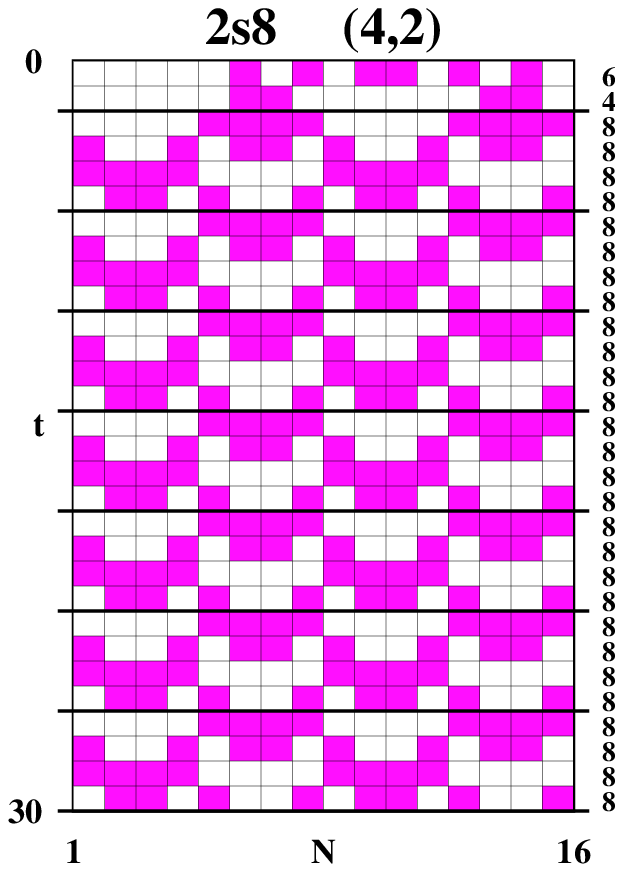}%
\includegraphics[width=4.75cm]{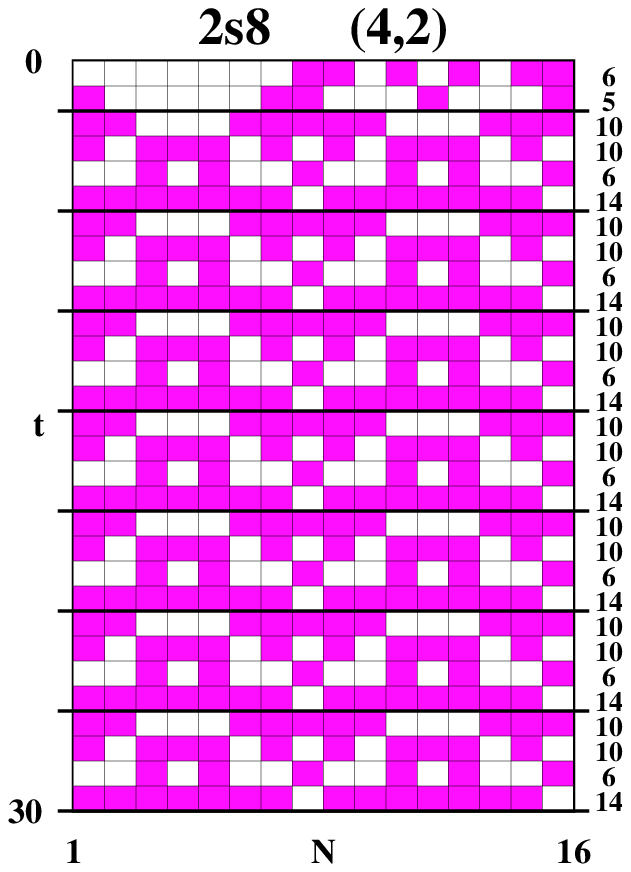}%
\includegraphics[width=4.75cm]{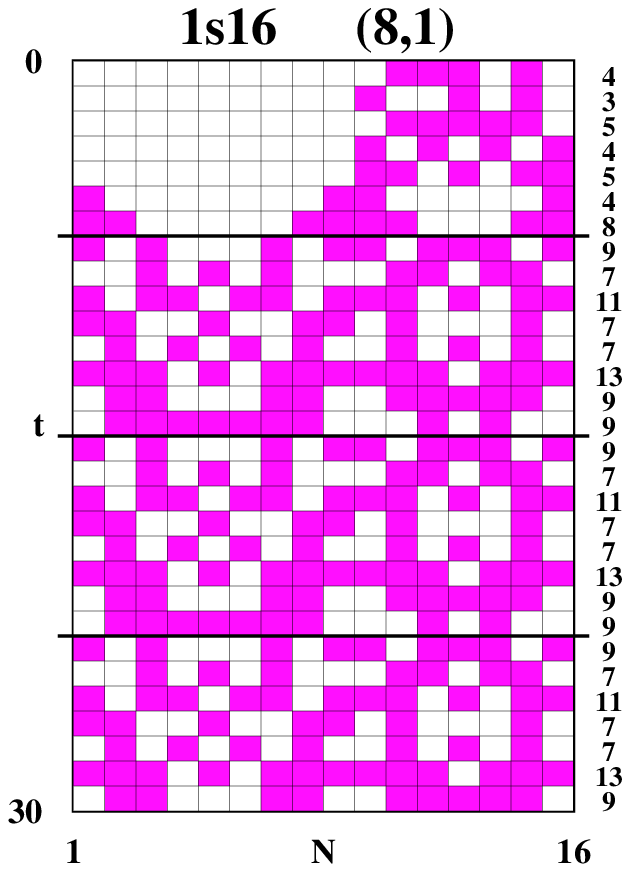}}
%\vspace{-0.4truecm}}
\end{center}
\caption{\protect Spatially primitive and doubling 
  patterns obtained for $N=16$ sites.
  The rightmost figure shows a 
  primitive pattern of spatial period 16,  
  denoted $1s16$, while the other two figures display 
  spatial-doublings, denoted $2s8$,
  of primitive patterns of spatial period 8.
  The notation  ``$(4,2)$'' indicates a pattern 4-periodic in time
  and 2-periodic in space.
  Vertical numbers on the right give the
  ``magnetization'', i.e.~the total amount of ``1'' at each
  time step.}
\label{fig:fig2}       
\end{figure}
%==================================================

The system is started at $t=0$ from a random initial configuration.
The leftmost portion of Fig.~\ref{fig:fig1} shows a
typical example of the
non-trivial behaviors studied by Wolfram, a so-called 
{\it glider} \cite{wolf84,ANKS,ila2001,bastien}, a localized structure that
propagates along the automaton.
The rightmost portion of Fig.~\ref{fig:fig1} shows an
example of the new behaviors reported here: the
spontaneous emergence of spatio-temporal order in the
automaton, following a transient ``equilibration'' time.
The great regularity seen in  the rightmost pattern in Fig.~\ref{fig:fig1}
contrasts sharply with ``more complex  than chaos" 
expectations for class 4.

Figure \ref{fig:fig2} shows some space-filling patterns
obtained for a lattice with $N=16$ sites. 
The two leftmost patterns  contain
spatial-doublings of two different primitive patterns observed for $N=8$.
Noteworthy is the fact that despite asymmetries in the
initial conditions, spatial periodicity sets in. 
On the rightmost pattern, notice the relatively long lack of
activity in half of the figure. 
% The vertical sequences of
% numbers indicate the ``magnetization'' (i.e.~the total
% quantity of ``1'' at each time step) as a function of time. 

%==================================================
\begin{figure}[bht]
\unitlength 1mm
\begin{center}
\leavevmode 
{\includegraphics[width=4.75cm]{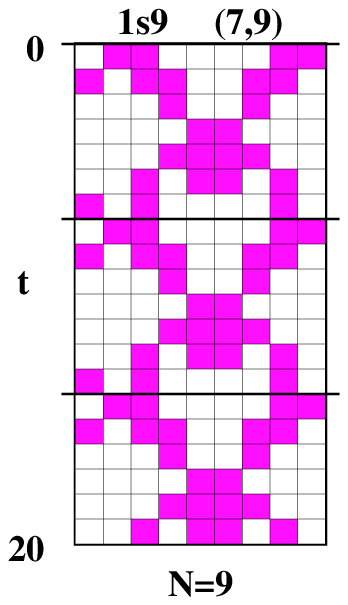}%
\includegraphics[width=4.75cm]{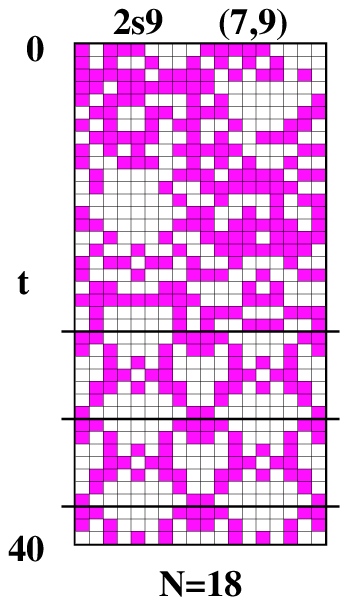}%
\includegraphics[width=4.75cm]{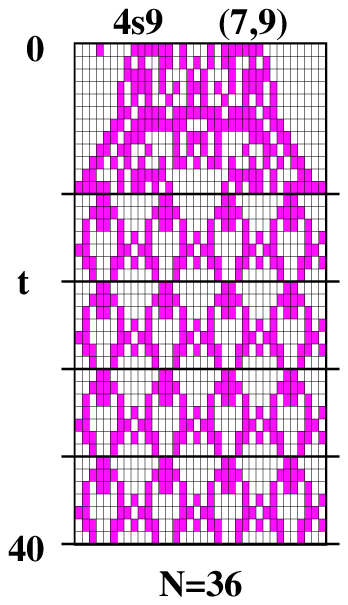}}
%\vspace{-0.4truecm}}
\end{center}
\caption{\protect Spatial doublings $\ell s9$, where
  $\ell=1$ indicates a {\it primitive\/} spatial period,
  $\ell=2$ indicates a doubling, etc.
  Here temporal and spatial periodicities produce
  a ``$(7,9)$-pattern'' namely, a pattern which is 7-periodic in time
  and 9-periodic in space.
  Initial conditions are generated randomly.}
\label{fig:fig3}       
\end{figure}
%==================================================

\begin{figure}
\unitlength 1mm
\begin{center}
\leavevmode 
%%% {\includegraphics[width=4.75cm,angle=90]{damage000.eps}}%
{\includegraphics[width=4.75cm]{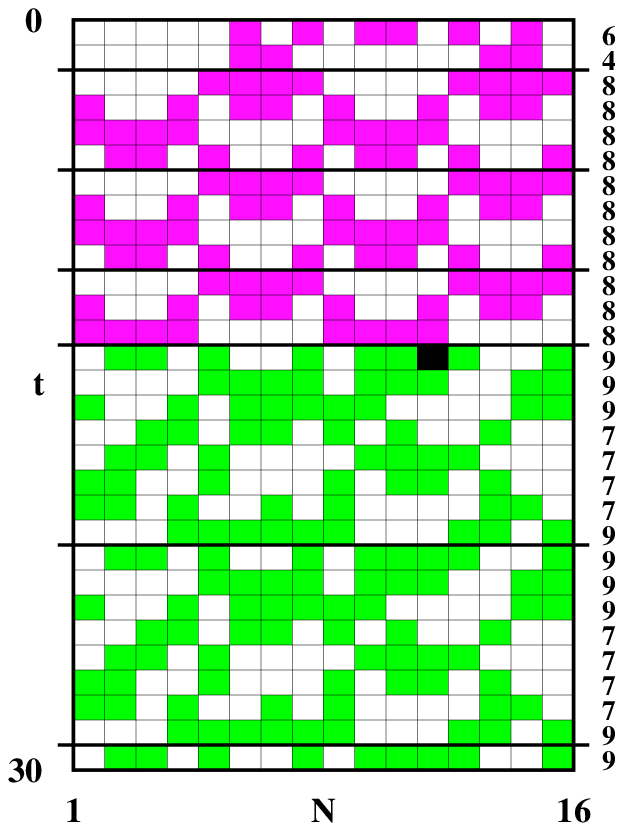}%
\includegraphics[width=4.75cm]{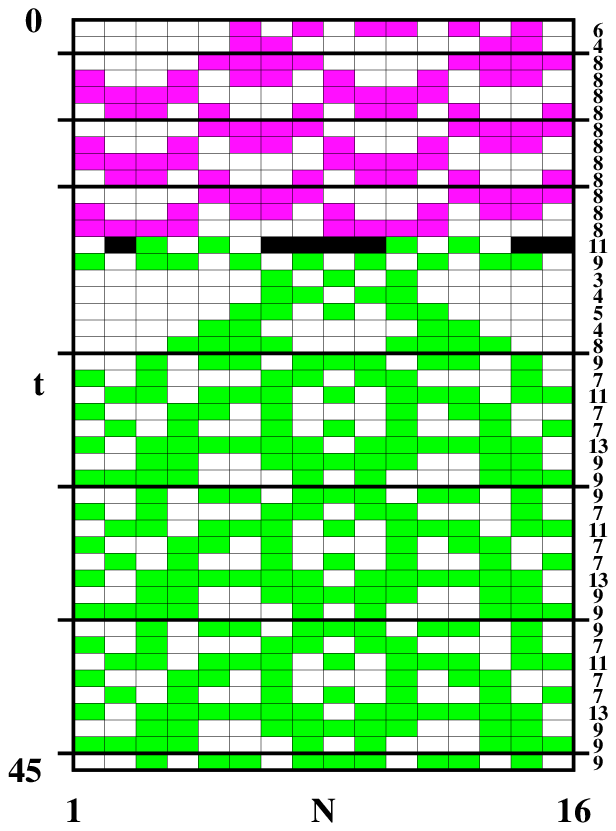}%
\includegraphics[width=4.75cm]{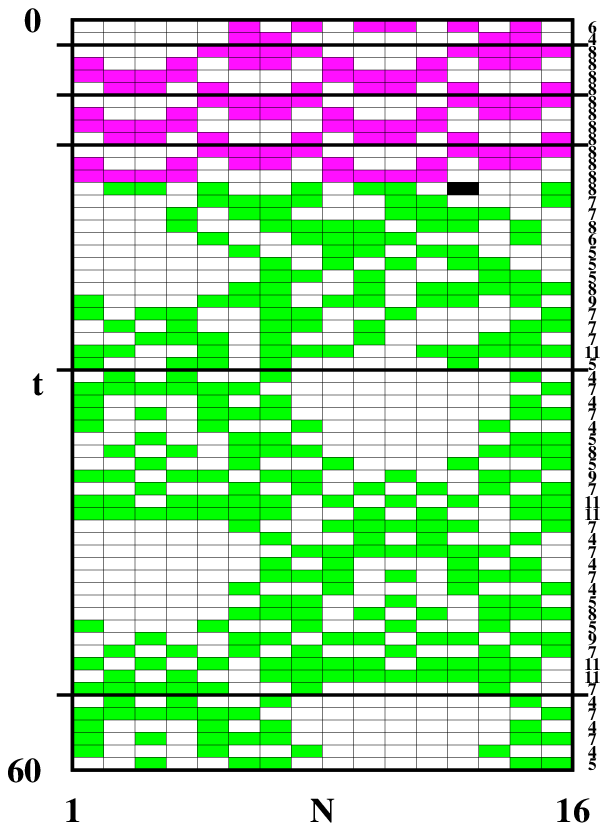}}%
%\vspace{-0.4truecm}}
\end{center}
\caption{\protect Effect of low and high damage (in black) in the 
         evolution of a given pattern.}
\label{fig:fig4}       
\end{figure}

\begin{figure}
\unitlength 1mm
\begin{center}
\leavevmode 
%%% {\includegraphics[width=4.75cm,angle=90]{damage000.eps}}%
{\includegraphics[width=3.25cm]{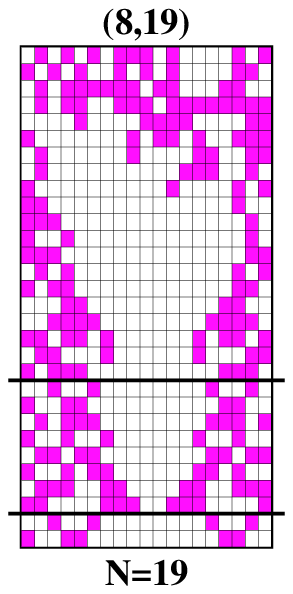}%
\includegraphics[width=3.25cm]{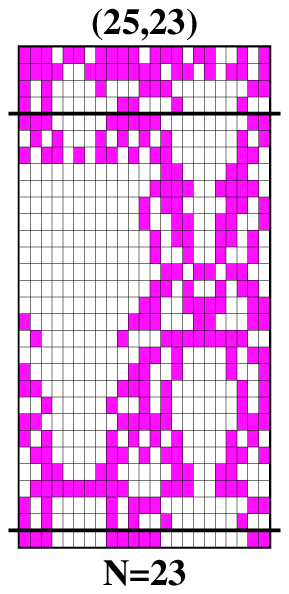}%
\includegraphics[width=3.25cm]{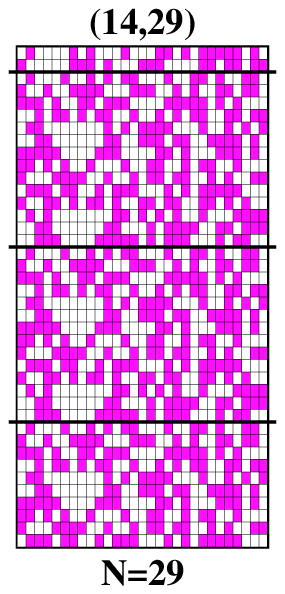}%
\includegraphics[width=3.25cm]{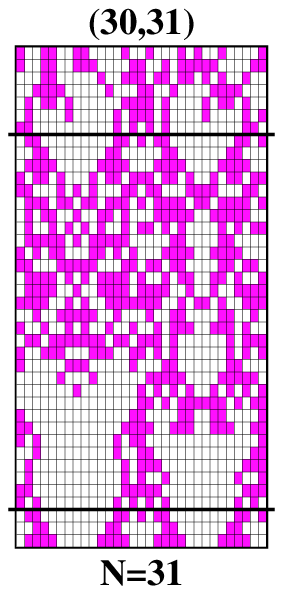}}%
%\vspace{-0.4truecm}}
\end{center}
\caption{\protect Primitive periods for odd lattice sizes, shown
  for 30 time steps. The structure seen for $N=19$
  grows until it touches itself and contracts, 
  pulsating periodically.}
\label{fig:fig5}       
\end{figure}

\begin{figure}
\unitlength 1mm
\begin{center}
\leavevmode 
%%% {\includegraphics[width=4.75cm,angle=90]{damage000.eps}}%
{\includegraphics[width=4.5cm]{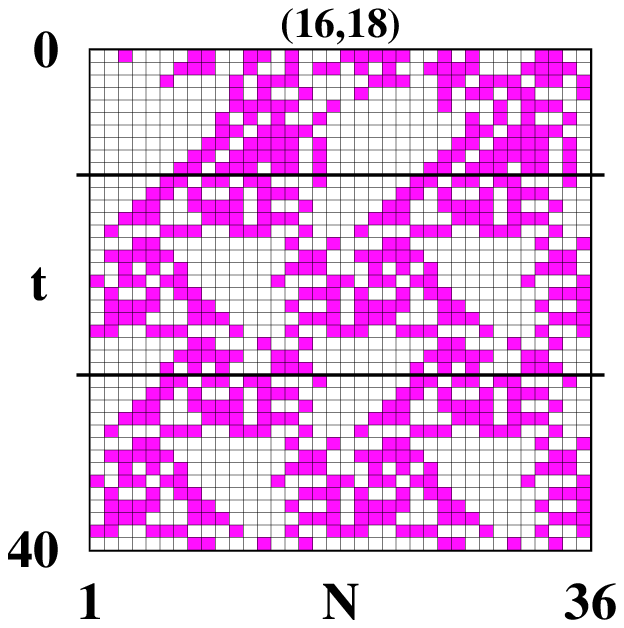}%
\includegraphics[width=4.5cm]{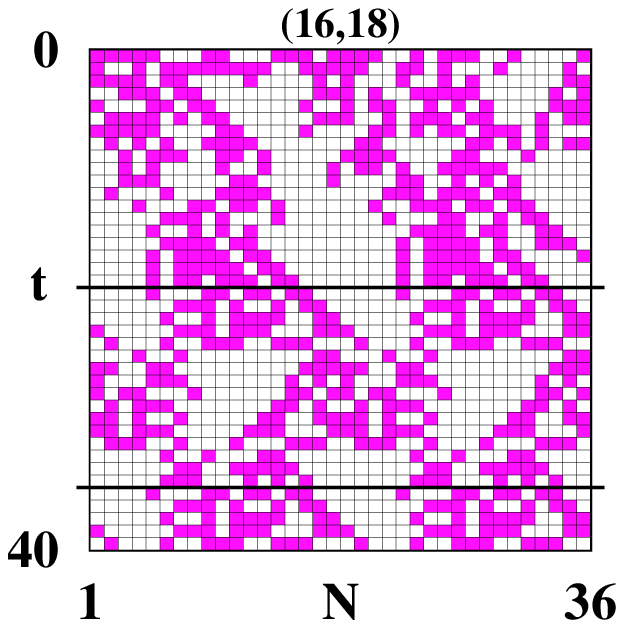}%
\includegraphics[width=4.5cm]{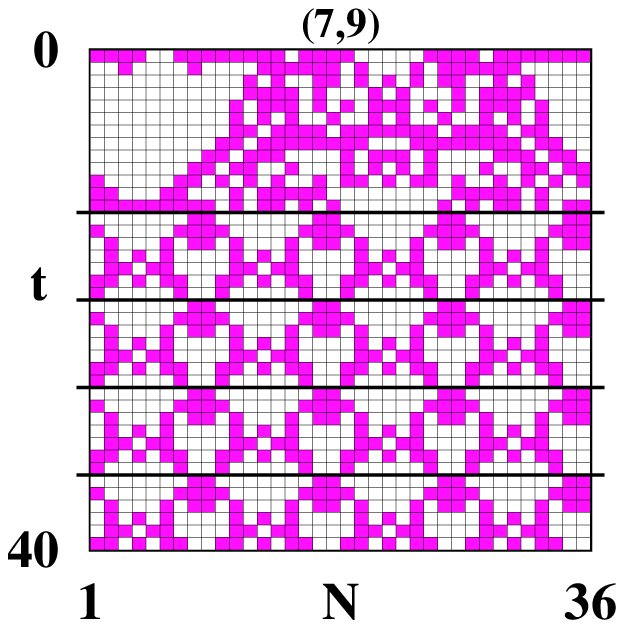}}%
%\vspace{-0.4truecm}}
\end{center}
\caption{\protect Different transients, similar starting phases
   and strongly correlated spatial patterns. 
   Compare the $(7,9)$ pattern here with the
   corresponding $4s9$ in Fig.~\ref{fig:fig3}. Note
   the common precursor ``gun'' emitting  periodic
   gliders. The precursor here has quite distinct initial
   conditions. Another similar  precursor may be seen in 
   Fig.~(\ref{fig:fig4}).}
\label{fig:fig6}       
\end{figure}

Figure \ref{fig:fig3} shows that it is possible to tile
periodically the space when $N$ grows. 
We stress that initial conditions are always random.
As the figure for $N=18$ shows, in some cases the transient
may be several times bigger than the final temporal periodicity
that sets in.
Figure \ref{fig:fig4} illustrates several interesting
things. It shows that spatial periodicity may be
easily induced by ``damaging'' the automata in randomly
chosen sites. The black cells indicate the time and position
of the damage, the actual damage consisting in reversing the
state of the site after having updated it using Eq.~(\ref{rule20}).
Figure \ref{fig:fig4} also shows that damage may not only 
induce transitions among
different patterns but is capable of inducing changes
in the spatial periodicity of the system, in particular a
frequently observed result is to see doublings of the
spatial period.
Figures \ref{fig:fig5} and \ref{fig:fig6} are explained in
their captions. 
We conclude recalling that small lattices and 
relatively short temporal evolutions like those reported here are
precisely the most interesting regimes for 
applications  \cite{ANKS,ila2001,bastien}. 
A promising possibility seems to be exploiting class 4
automata as a tool to compress digital sound and images.
Detailed results will be reported elsewhere.

%==================================================
%==================================================

%==================================================
%==================================================

%==================================================
%==================================================

\vspace{0.5truecm}

%%%%%%%%%%%%%%%%%%%%%%%%%%%\begin{acknowledgments}
JACG thanks Sonderforschungsbereich 404, 
Germany, and CNPq, Brazil, for financial support.
%This work was also supported by Project 077/2001 
%sponsored by CAPES (Brazil) and GRICES (Portugal).
HJH thanks the Max-Planck-Forschungspreis awarded to him.
%%%%%%%%%%%%%%%%%%%%%%%%%%%\end{acknowledgments}

\end{document}